\documentclass[prd,aps,floats,floatfix,amsmath,nofootinbib,superscriptaddress]{revtex4}
\usepackage{setspace}
\usepackage{axodraw4j}
\usepackage{pstricks}
\usepackage{graphicx}
\usepackage{bm}
\usepackage{epsfig}
\usepackage{latexsym}
\usepackage{color}
\usepackage{colordvi}

\newcommand{\nn}{\nonumber}

\newcommand{\beq}{\begin{equation}}
\newcommand{\eeq}{\end{equation}}
\newcommand{\bea}{\begin{eqnarray}}
\newcommand{\eea}{\end{eqnarray}}
\begin{document}

\title{Viscosity Sum Rules at Large Scattering Lengths}

\author{Walter D. Goldberger and Zuhair U. Khandker}
\affiliation{
Department of Physics, Yale University, New Haven, CT 06520}

\begin{abstract}
We use the operator product expansion (OPE) and dispersion relations to obtain new model-independent ``Borel-resummed" sum rules for both shear and bulk viscosity of many-body systems of spin-1/2 fermions  with predominantly short range $S$-wave interactions.  These sum rules relate Gaussian weights of the frequency-dependent viscosities to the Tan contact parameter ${\cal C}(a)$.   Our results are valid for arbitrary values of the scattering length $a$, but receive small corrections from operators of dimension $\Delta>5$ in the OPE, and can be used to study transport properties in the vicinity of the $a\rightarrow\infty$ fixed point.  In particular, we find that the \emph{exact} dependence of the shear viscosity sum rule on scattering length is controlled by the function ${\cal C}(a)$.  The sum rules that we obtain depend on a frequency scale $\omega_0$ that can be optimized to maximize their overlap with low-energy data.
\end{abstract}

\maketitle

\section{\label{Introduction} Introduction}

Fermions with tuned short-range interactions, resulting in large $S$-wave scattering length, seem to describe a wide range of physical phenomena arising in diverse areas of physics, from nuclear and hadronic processes to table top experiments in condensed matter and atomic physics.   This is because the unitary limit, in which the scattering length $a$ diverges, is governed by an ultraviolet fixed point of the renormalization group in which the dynamics becomes universal. Progress over the last few years in manipulating samples of cold atoms tuned to a Feshbach resonance has made it possible to study the many-body dynamics of these systems in a detailed and controlled manner.

One of the ways in which universality manifests itself in these systems was uncovered in the work of S. Tan~\cite{tan1,tan2,tan3}, which derives a  set of relations that hold true when evaluated in any quantum state, and which are fixed, for $a\rightarrow\infty$, in terms of  few-body dynamics at short length scales.   It was subsequently pointed out by Braaten and Platter~\cite{bp} that the Tan relations can be derived from the operator product expansion (OPE)~\cite{wilson} applied to the quantum field theory realization of fermions near unitarity,
\begin{equation}
\mathcal{L} = \sum_{\alpha=1,2} \psi^\dagger_\alpha \left(i\partial_t + \frac{\nabla^2}{2m}\right)\psi_\alpha + g \psi^\dagger_1 \psi^\dagger_2 \psi_1 \psi_2.
\label{L}
\end{equation}
Indeed, the existence of such universal relations in field theory is natural from the point of view of the existence of an ultraviolet fixed point $a\rightarrow\infty$ with enhanced non-relativistic conformal (Schrodinger) symmetry~\cite{Hagen:1972pd, niederer,Mehen:1999nd,Nishida:2007pj}.    Since the original work of~\cite{tan1,tan2,tan3}, a large class of universal relations have been obtained using OPE methods~\cite{Braaten:2010dv,st,GnR,barth} or otherwise~\cite{rf,photo,legget,stat,TR,EHZ}.  See~\cite{Braaten:2010if} for a recent review of these developments.

In this paper, we report on a new set of sum rules that relate weighted integrals of the frequency-dependent shear and bulk viscosities $\mbox{Re}\,\eta(\omega)$, $\mbox{Re}\,\zeta(\omega)$ of a many-body system near unitarity to the contact parameter ${\cal C}$ introduced in~\cite{tan1}.   In order to obtain these sum rules, we first use the OPE to fix the $\omega\rightarrow\infty$ asymptotic behavior of the Feynman (time ordered) two-point correlator of the traceless part of the stress tensor of the theory defined by Eq.~(\ref{L}). (The analogous OPE for the particle number density is known~\cite{GnR}, and that of the particle number current is presented in the appendix as an independent check of results.) By standard Kubo formulae, together with analyticity, these OPEs give the tail behavior of the viscosities at $\omega\rightarrow\infty$.  The result for the tail of $\mbox{Re}\,\eta(\omega)$ agrees with the previously obtained expression in ref.~\cite{EHZ}, while the asymptotic bulk viscosity is new (see also refs.~\cite{TR,Chao:2010tk} for previous work on the viscosity of unitary gases).  Unlike ref.~\cite{EHZ}, however, our expressions are valid for arbitrary values of the parameter $a\sqrt{m\omega}$.   They are accurate up to calculable corrections to the OPE from operators of dimension $\Delta>5$ (e.g., three-body contact operators as well as finite range effects).    Finally, in order to obtain results that can be compared to low-energy experiments, we use the methods introduced in ref.~\cite{GnR} (themselves based on the QCD sum rule technology of~\cite{SVZ}), which employ a Borel resummation of the OPE to derive sum rules that can be made to overlap with the data by adjusting a free parameter $\omega_0$.

This paper is organized as follows.  Sec.~\ref{Preliminaries} collects several previous results that will be used in our derivation of sum rules, including the relevant Kubo formulae in sec.~\ref{Kubo Formulas} and a review of the interplay between dispersion relations and the OPE in sec.~\ref{Sum Rules and the OPE}.   In sec.~\ref{TTOPE} we compute the leading term for the OPE of the traceless-part of the stress tensor with itself, working to all orders in the coupling constant $g$ of Eq.~(\ref{L}) but to leading order (dimension $\Delta=4$) in the operator expansion.   This, together with the previous results of~\cite{GnR} on the structure factor $S(\omega,{\vec q})$ allows us to obtain in sec.~\ref{Results} the asymptotic properties of both bulk and shear viscosity as well as sum rules to all orders in $a$.    The scattering length dependence of the shear viscosity results is given exactly by that of the contact ${\cal C}(a)$, while the expressions involving bulk viscosity vanish as $a^{-2}$ in the limit of infinite scattering length.  In appendix~\ref{ope} we give the full OPE for the longitudinal and transverse parts of the current two-point function, which gives an independent derivation of our results and serves as a consistency check.   We conclude in sec.~\ref{Conclusion}

\section{Preliminaries}
\label{Preliminaries}

\subsection{\label{Kubo Formulas} Hydrodynamic Relations}

The connection between hydrodynamics and the microscopic theory of Eq.~(\ref{L}) is given by standard Kubo formulae.    We will focus on viscosity, which at the macroscopic level is defined in terms of a small gradient expansion of the stress tensor $T^{ij}(x)$ :
\begin{equation}
T_{ij}(x) = P(x)\delta_{ij} + \rho(x) u_i(x) u_j(x) - \sigma_{ij}(x),
\end{equation}      
where $x=(t,{\vec x})$.  In this equation $P(x)$ is the pressure, $\rho(x)$ is mass density,  and $\vec{u}(x)$ is the fluid velocity.    The dissipative part of the stress tensor, in the large wavelength limit, is given by
\begin{equation}
\sigma_{ij} = \eta\left[\partial_i u_j + \partial_j u_i - \frac{2}{d}\delta_{ij}(\nabla\cdot u)\right] + \zeta \delta_{ij} (\nabla\cdot u) + \cdots,
\end{equation}           	
where $d$ is the number of spatial dimensions and $\cdots$ denotes higher-order gradients of $\vec{u}$.    Thus the long wavelength fluid transport properties are encoded in terms of two transport coefficients, namely the shear viscosity $\eta$ and bulk viscosity $\zeta$.

At the microscopic level, $\eta$ and $\zeta$ are the zero-frequency limits of particular linear response functions associated with the quantum theory.   Specifically, for a non-relativistic theory with conserved particle number $J^\mu=(n(x),{\vec J}(x))$, consider the response function
\begin{equation}
\label{eq:chiR}
\chi_R^{i,j} (q) = i\int \mathrm{d}^{d+1}x\, e^{iq\cdot x}\, \theta(t) \left\langle \left[J^{i}(x),J^{j}(0)\right]\right\rangle,
\end{equation}
with the notation $x^\mu=(t,{\vec x})$, $q^\mu=(\omega,{\vec q})$, $q\cdot x = \omega t - {\vec q}\cdot {\vec x}$, and $d^{d+1} x=dt d^d {\vec x}$.  Decomposing this Green's function into longitudinal and transverse components
\begin{equation}
\chi_R^{i,j}(q) = \chi_L(q) \frac{q^i q^j}{\vec{q}^2} + \chi_T(q) \left(\delta^{ij} - \frac{q^i q^j}{\vec{q}^2} \right),
\end{equation}
one defines~\cite{Forster} frequency dependent viscosities $\eta(\omega)$ and $\zeta(\omega)$ by
\begin{eqnarray}
\eta(\omega) &=& \displaystyle\lim_{\vec{q}\to 0} \frac{-i\omega m^2}{\vec{q}^2} \chi_T (\omega, \vec{q}) \label{Kubo1},\\
\zeta(\omega) + \frac{2d-2}{d}\eta(\omega) &=& \displaystyle\lim_{\vec{q}\to 0} \frac{-i\omega m^2}{\vec{q}^2} \chi_L (\omega, \vec{q}).
\end{eqnarray}
With these conventions, the dissipative response is encoded in the \emph{real} parts of the complex functions 
$\eta(\omega)$ and $\zeta(\omega)$,
\begin{eqnarray}
\mathrm{Re}\,\eta(\omega) &=& \displaystyle\lim_{\vec{q}\to 0} \frac{m^2 \omega}{\vec{q}^2} \mathrm{Im}\, \chi_T (\omega, \vec{q}) \label{Kubo1.5},\\
\mathrm{Re}\,\zeta(\omega) + \frac{2d-2}{d}\mathrm{Re}\,\eta(\omega) &= &\displaystyle\lim_{\vec{q}\to 0} \frac{m^2 \omega}{\vec{q}^2} \mathrm{Im}\, \chi_L (\omega, \vec{q}) \label{Kubo2}.
\end{eqnarray}
The constants $\eta$ and $\zeta$ are the zero-frequency limits of these functions: $\eta \equiv \mathrm{Re}\,\eta(\omega =0)$, $\zeta \equiv \mathrm{Re}\,\zeta(\omega =0)$.

In addition to the number density and current $(n,{\vec J}(x))$, hydrodynamic transport is characterized by several other conserved  quantities, namely energy density $T^{00}={\cal H}$, energy flux $T^{i0}$ and stress $T^{ij}$.    These obey continuity equations
\begin{eqnarray}
\partial_t n + \partial_i J^i = 0, \label{Consv3}\\
\partial_t T^{00} + \partial_i T^{i0} = 0, \label{Consv1}\\
m \partial_t J^i + \partial_j T^{ji} = 0. \label{Consv2}
\end{eqnarray}
As a result $\mathrm{Re}\,\eta(\omega)$ and $\mathrm{Re}\,\zeta(\omega)$ can also be obtained from the stress tensor correlator
\begin{equation}
\chi_R^{ij,k\ell} (q) = i\int \mathrm{d}^{d+1}x\, e^{iq\cdot x}\, \theta(t) \left\langle \left[T^{ij}(x),T^{k\ell}(0)\right]\right\rangle.
\end{equation}
Thus we can write,
\begin{align}
\mathrm{Re}\,\eta(\omega) &= \displaystyle\lim_{\vec{q}\to 0} \frac{1}{\omega} \mathrm{Im}\, \chi_R^{xy,xy} (\omega, \vec{q}) ,\label{Kubo3}\\
\mathrm{Re}\,\zeta(\omega) + \frac{2d-2}{d}\mathrm{Re}\,\eta(\omega) &= \displaystyle\lim_{\vec{q}\to 0} \frac{1}{\omega} \mathrm{Im}\, \chi_R^{xx,xx} (\omega, \vec{q}) \label{Kubo4}.
\end{align}
Additionally, conservation of number current relates the longitudinal function $\chi_L(\omega,{\vec q})$ to the correlator of the number density\footnote{This relation holds modulo Schwinger terms which are analytic in momentum space and therefore do not contribute to dissipative response.}
\begin{equation}
\chi_L(\omega,{\vec q}) = {i\omega^2\over {\vec q}^2} \int d^{d+1} x e^{iq\cdot x} \theta(t)\langle [n(x),n(0)]\rangle,
\end{equation}
and therefore 
\begin{equation}
\label{eq:K5}
\mathrm{Re}\,\zeta(\omega) + \frac{2d-2}{d}\mathrm{Re}\,\eta(\omega) = \displaystyle\lim_{\vec{q}\to 0} {\pi m^2\omega^3\over {\vec q}^4} S(\omega,{\vec q}),
\end{equation}
where $S(\omega,{\vec q})$ is the dynamic structure factor.

In this paper we will find it most convenient to use Eqns.~(\ref{Kubo3}),~(\ref{eq:K5}) to obtain viscosity sum rules.  In order to obtain hydrodynamic transport coefficients for the theory defined by Eq.~(\ref{L}) we need operator expressions for the conserved currents.    The particle number density and current operators are
\begin{eqnarray}
n(x) &=& \psi^\dagger\psi(x),\\
\vec J(x) &=& -{i\over 2m} \psi^\dagger \stackrel{\leftrightarrow}{\nabla}\psi.
\end{eqnarray}
For the purposes of this paper, it is sufficient to define the stress-tensor as the spatial part of the Noether energy-momentum tensor $T^{\mu\nu}$ associated with invariance under time plus space translations, whose components are~\cite{Jackiw:1990mb} energy density,
\begin{equation}
\mathcal{H} = T^{00} = \frac{1}{2m}\left|\nabla\psi\right|^2 - g \psi^\dagger_1 \psi^\dagger_2 \psi_1 \psi_2,
\end{equation}
energy flux,
\begin{equation}
T^{i0} = -\frac{1}{2m}\left[ \left(\partial_t \psi^\dagger\right)\left(\partial_i \psi\right) +  \left(\partial_i \psi^\dagger\right)\left(\partial_t \psi\right) \right], 
\end{equation}
and stress tensor
\begin{equation}
\label{stress}
T^{ij} = \frac{1}{2m}\left[ \left(\partial_i \psi^\dagger\right)\left(\partial_j \psi\right) +  \left(\partial_j \psi^\dagger\right)\left(\partial_i \psi\right) \right]-\delta^{ij}\left[{1\over 4m} \nabla^2(\psi^\dagger\psi) + g \psi_1^\dagger \psi_2^\dagger \psi_1\psi_2\right].
\end{equation}
The remaining component is the momentum density, which obeys $T^{0i}= m J^i$ as required by Galilean invariance.

 \subsection{\label{Sum Rules and the OPE} OPE and Sum Rules}

The unitary limit is characterized by strong coupling, and thus the problem of computing the functions $\eta(\omega)$ and $\zeta(\omega)$ in the many-body ground state is generally not analytically tractable.    However, as discussed in refs.~\cite{bp,Braaten:2010dv,st,GnR}, it is possible to obtain some analytic results in the $\omega\rightarrow\infty$ limit.   This is because in the limit $q^\mu\rightarrow\infty$, the product of operators appearing in two-point correlators can be expanded, via the OPE, in a set of local operators with coefficients that are calculable in the few-body sector, in which the theory is integrable, without reference to the many-body ground state. While this limit is not directly related to hydrodynamic response at large length and time scales, it can be connected to hydrodynamics by employing dispersion relations, as was done in ref.~\cite{GnR}.  This gives model-independent predictions in the form of sum rules.   In this section we briefly summarize the procedure for obtaining such sum rules from the OPE and refer to~\cite{GnR} for details.

Starting from the Kramers-Kronig dispersion relations for a causal Green's functions $\chi(\omega)$ (such as the functions $\chi_{L,T}(\omega,{\vec q})$ introduced the previous section),
\bea
\label{eq:disp}
\nonumber
\mbox{Re }\chi(\omega) &=& \mbox{Pr } \int_0^\infty {d\omega'^2\over \pi} {\mbox{Im } \chi(\omega')\over {\omega'}^2 - \omega^2},\\
\mbox{Im } \chi(\omega) &=& -2 \omega \mbox{Pr } \int_0^\infty {d\omega'\over \pi} {\mbox{Re } \chi(\omega')\over {\omega'}^2 - \omega^2},
\eea
where $\mbox{Pr}$ is the Cauchy principal part, one obtains upon taking the formal limit $\omega\rightarrow\infty,$ 
\beq
\label{eq:naive}
\mbox{Re } \chi(\omega\rightarrow\infty) =  -\sum_{n=0}^\infty {1\over (\omega^2)^{n+1}} \int_0^\infty {d\omega'^2\over \pi} {\omega'}^{2n} {\mbox{Im }\chi(\omega')}.
\eeq
Therefore, frequency averages of the dissipative function $\mbox{Im}\chi(\omega)$ are determined by the high-frequency asymptotic behavior of the Green's function.   In field theory, $\chi(\omega)$ generically has power-law ultraviolet behavior and its is not valid to expand the dispersion relations.   This can be circumvented~\cite{SVZ} by employing a Borel resummation of the dispersion relation, which gives
\beq
\label{eq:borelsr}
{1\over \omega_0^2}\int_0^\infty {d\omega^2 \over\pi} e^{-\omega^2/\omega_0^2} \mbox{Im} \chi(\omega) = \sum_\alpha {a_\alpha\over\Gamma(\alpha)} (\omega_0^2)^{-\alpha},
\eeq
where the expansion coefficients $a_\alpha$ are fixed by the asymptotic behavior of the Green's function at large imaginary frequencies\begin{equation}
\label{eq:as}
\chi(i\omega\rightarrow\infty)\sim\sum_\alpha a_\alpha (\omega^2)^{-\alpha}.
\end{equation}
Here we have used the property that  $\chi(i\omega)$ is a real function for real $\omega>0$~\cite{LL}.   If the free parameter $\omega_0$ is taken sufficiently large relative to the scales characterizing the ground state, the sum on the RHS of Eq.~(\ref{eq:borelsr}) can be truncated after the first few terms.    It becomes a phenomenological/experimental problem to determine an optimal value of $\omega_0$ for which the integral on the LHS of Eq.~(\ref{eq:borelsr}) is dominated by the low energy data while retaining convergence on the RHS.  Note that in general, the series is asymptotic although for conformal theories (such as the theory of fermions at unitarity) the radius of convergence is finite.

At the linear response level, the function $\chi(\omega)$ can be expressed as a two-point correlation function of operators.    Thus the coefficients appearing in the expansion of Eq.~(\ref{eq:borelsr}) can be understood using the operator product expansion (OPE)~\cite{bp,Braaten:2010dv,st,GnR}.   For a non-relativistic QFT, the OPE is the statement that a product of local operators has the following expansion in the limit of zero separation,
\begin{equation}
A(x)B(0) \underset{x\to 0}{\sim} \sum_{\alpha} \left|\vec{x}\right|^{\Delta_\alpha-\Delta_A-\Delta_B} f_\alpha\left(\frac{\left|\vec{x}\right|^2}{t}\right) \mathcal{O}_\alpha(0).
\label{NROPE}
\end{equation}
Eq.~(\ref{NROPE}) is an operator statement which holds inside arbitrary matrix elements.    The sum on the right hand side is over operators with scaling dimension $\Delta_\alpha$, and  the Wilson coefficients $ f_\alpha\left({\left|\vec{x}\right|^2/t}\right)$ are calculable by taking few-particle matrix elements on both sides of the OPE relation\footnote{We are simplifying the discussion slightly by ignoring possible renormalization group scale dependence of the Wilson coefficients (in even spatial dimensions), or dependence on scattering length away from the fixed point $a\rightarrow\infty$.   However, our results in the next section take account of the exact $a$-dependence of the OPE.}.

Inserting Eq.~(\ref{NROPE}) inside a time-ordered Green's functions gives the $q^\mu\rightarrow\infty$ asymptotic behavior 
\begin{equation}
\label{eq:GOPE}
G^{AB}(q) = i\int d^{d+1} x e^{iq\cdot x} \langle T A(x) B(0)\rangle \sim \sum_{\alpha} \frac{1}{\omega^{\left[(d+2)+\Delta_\alpha-\Delta_A-\Delta_B\right]/2}} c_\alpha\left(\frac{\vec{q}^2}{2m\omega}\right) \left\langle \mathcal{O}_\alpha(0) \right\rangle
\end{equation}
In the $\omega\rightarrow\infty,$ fixed ${\vec q}$ limit considered in this paper, the Wilson coefficients $c_\alpha(z)$ are analytic functions away from multi-particle thresholds.    The Feynman Green's function $G^{AB}(q)$ agrees with the causal correlator away from real frequencies, and therefore the OPE reproduces the expansion in Eq.~(\ref{eq:as}), relating the coefficients $a_\alpha$ to the condensates $\langle O_\alpha\rangle$.    These condensates are functions of thermodynamic variables that characterize the many-body ground state and thus scale as powers $\langle O_\alpha\rangle\sim \mu^{\Delta_\alpha/2}$ of the Fermi energy.   The OPE is valid for $\omega\gg\mu$ (but smaller than the energy scale associated with finite effective range corrections to Eq.~(\ref{L})).

\section{\label{Shear viscosity sum rule}Viscosity sum rules}
\subsection{OPE}
\label{TTOPE}
We now determine the asymptotic $\omega\rightarrow\infty$ behavior of the frequency-dependent viscosities $\eta(\omega),\zeta(\omega)$.   In order to do so, we construct the OPE for the product
\begin{equation}
M^{ij,kl}(\omega,\vec{q}=0) \equiv \int \mathrm{d}^{d+1}x\, e^{iq\cdot x}\, T\{\tau^{ij}(x)\tau^{kl}(0)\},
\label{StressOPE}
\end{equation}
where $\tau_{ij}(x)$ is the traceless part of the stress tensor in Eq.~(\ref{stress}), $\tau_{ij}=T_{ij} -\delta_{ij} T_{kk}/d$.    For our purposes it is sufficient to take $q^\mu=(\omega\rightarrow\infty,{\vec q}\rightarrow 0)$.

The OPE is constructed by taking matrix elements of Eq.~(\ref{StressOPE}) between few-particle states.  In the limit $q^\mu\rightarrow\infty$ these matrix elements can be expressed by local operator matrix elements, with $q$ dependent coefficients that we calculate below.     At leading order in the OPE expansion, the necessary matrix elements are between one-particle and two-particle states.   We can take the limit in which these external states have vanishing energy and momentum.   This omits the contributions from local operators constructed from at least one pair of Fermi fields $(\psi(x),\psi^\dagger(x))$ and at least two derivatives (by rotational invariance, operators with one derivative will ultimately give vanishing contributions to Green's functions in the many-body ground state).  

\begin{figure*}[t]
\begin{center}
\fcolorbox{white}{white}{
  \begin{picture}(423,90) (127,-168)
    \SetWidth{1.0}
    \SetColor{Black}
    \Line[arrow,arrowpos=0.5,arrowlength=5,arrowwidth=2,arrowinset=0.2](169,-112)(265,-112)
    \Line[arrow,arrowpos=0.5,arrowlength=5,arrowwidth=2,arrowinset=0.2](508,-112)(549,-167)
    \Line[arrow,arrowpos=0.5,arrowlength=5,arrowwidth=2,arrowinset=0.2](413,-112)(509,-112)
    \Line[arrow,arrowpos=0.5,arrowlength=5,arrowwidth=2,arrowinset=0.2](128,-166)(169,-111)
    \Line[arrow,arrowpos=0.5,arrowlength=5,arrowwidth=2,arrowinset=0.2](264,-112)(305,-167)
    \COval(266,-113)(7.071,7.071)(45.0){Black}{White}\Line(262.464,-116.536)(269.536,-109.464)\Line(262.464,-109.464)(269.536,-116.536)
    \COval(168,-113)(7.071,7.071)(45.0){Black}{White}\Line(164.464,-116.536)(171.536,-109.464)\Line(164.464,-109.464)(171.536,-116.536)
    \Text(334,-113)[lb]{\Large{\Black{$+$}}}
    \Text(155,-99)[lb]{\Large{\Black{$q\uparrow$}}}
    \Text(497,-99)[lb]{\Large{\Black{$q\uparrow$}}}
    \Line[arrow,arrowpos=0.5,arrowlength=5,arrowwidth=2,arrowinset=0.2](372,-166)(413,-111)
    \COval(413,-112)(7.071,7.071)(45.0){Black}{White}\Line(409.464,-115.536)(416.536,-108.464)\Line(409.464,-108.464)(416.536,-115.536)
    \COval(508,-112)(7.071,7.071)(45.0){Black}{White}\Line(504.464,-115.536)(511.536,-108.464)\Line(504.464,-108.464)(511.536,-115.536)
  \end{picture}
  }
\caption{Diagrams contributing to the matrix element  of the operator $M^{ij,kl}(q)$ between one-particle states.   The symbol $\otimes$ denotes an insertion of the operator $\tau^{ij}$. \label{fig:1to1}}
\end{center}
\end{figure*}
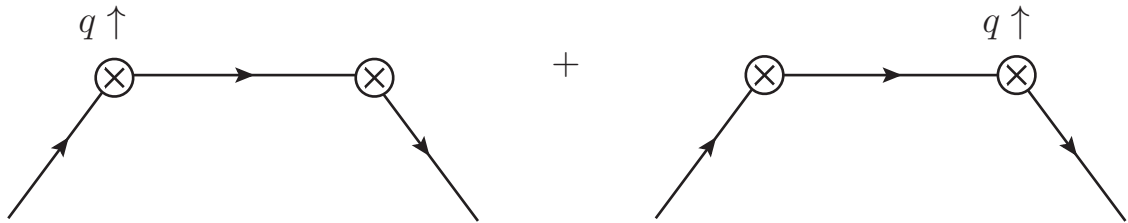

The one-particle matrix element of $M^{ij,kl}(q)$ is given by the graphs in Fig.~\ref{fig:1to1}.   These are exact to all orders in $g$.   Given that 
\begin{equation}
\label{eq:trule}
\langle p'|\tau_{ij}(q)|p\rangle = (2\pi)^{d+1}\delta(p-q-p')\cdot {1\over 2m} \left[{\vec p}_i ({\vec p}-{\vec q})_j + {\vec p}_j ({\vec p}-{\vec q})_i -{2\over d} \delta_{ij} {\vec p}\cdot ({\vec p}-{\vec q})\right],
\end{equation}
we see that both diagrams are zero in the limit of zero external momentum.
Consequently, there is no contribution from operators constructed from one pair $(\psi(x),\psi^\dagger(x))$ to the asymptotic behavior of the frequency dependent shear viscosity.

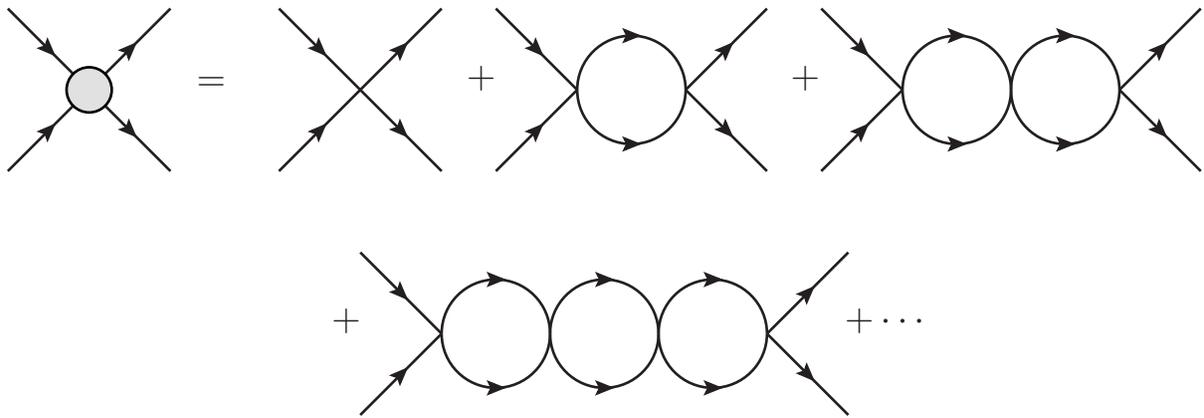
\begin{figure*}[t!]
\begin{center}
\fcolorbox{white}{white}{
  \begin{picture}(451,154) (26,-26)
    \SetWidth{0.5}
    \SetColor{Black}
    \Text(98.77,97.067)[lb]{\Large{\Black{$=$}}}
    \Text(200.945,97.067)[lb]{\Large{\Black{$+$}}}
    \Text(323.556,97.067)[lb]{\Large{\Black{$+$}}}
    \Text(343.991,5.109)[lb]{\Large{\Black{$+\cdots$}}}
    \SetWidth{1.0}
    \Line[arrow,arrowpos=0.5,arrowlength=5,arrowwidth=2,arrowinset=0.2](27.247,127.72)(57.9,97.067)
    \Line[arrow,arrowpos=0.5,arrowlength=5,arrowwidth=2,arrowinset=0.2](27.247,66.414)(57.9,97.067)
    \Line[arrow,arrowpos=0.5,arrowlength=5,arrowwidth=2,arrowinset=0.2](57.9,97.067)(88.552,66.414)
    \Line[arrow,arrowpos=0.5,arrowlength=5,arrowwidth=2,arrowinset=0.2](57.9,97.067)(88.552,127.72)
    \GOval(57.9,97.067)(8.515,8.515)(0){0.882}
    \Line[arrow,arrowpos=0.5,arrowlength=5,arrowwidth=2,arrowinset=0.2](129.422,127.72)(160.075,97.067)
    \Line[arrow,arrowpos=0.5,arrowlength=5,arrowwidth=2,arrowinset=0.2](129.422,66.414)(160.075,97.067)
    \Line[arrow,arrowpos=0.5,arrowlength=5,arrowwidth=2,arrowinset=0.2](160.075,97.067)(190.728,127.72)
    \Line[arrow,arrowpos=0.5,arrowlength=5,arrowwidth=2,arrowinset=0.2](160.075,97.067)(190.728,66.414)
    \Line[arrow,arrowpos=0.5,arrowlength=5,arrowwidth=2,arrowinset=0.2](211.163,127.72)(241.816,97.067)
    \Line[arrow,arrowpos=0.5,arrowlength=5,arrowwidth=2,arrowinset=0.2](211.163,66.414)(241.816,97.067)
    \Arc[arrow,arrowpos=0.5,arrowlength=5,arrowwidth=2,arrowinset=0.2,clock](262.251,97.067)(20.435,-180,-360)
    \Arc[arrow,arrowpos=0.5,arrowlength=5,arrowwidth=2,arrowinset=0.2](262.251,97.067)(20.435,-180,0)
    \Line[arrow,arrowpos=0.5,arrowlength=5,arrowwidth=2,arrowinset=0.2](282.686,97.067)(313.339,66.414)
    \Line[arrow,arrowpos=0.5,arrowlength=5,arrowwidth=2,arrowinset=0.2](282.686,97.067)(313.339,127.72)
    \Line[arrow,arrowpos=0.5,arrowlength=5,arrowwidth=2,arrowinset=0.2](333.774,66.414)(364.426,97.067)
    \Line[arrow,arrowpos=0.5,arrowlength=5,arrowwidth=2,arrowinset=0.2](333.774,127.72)(364.426,97.067)
    \Arc[arrow,arrowpos=0.5,arrowlength=5,arrowwidth=2,arrowinset=0.2,clock](384.861,97.067)(20.435,-180,-360)
    \Arc[arrow,arrowpos=0.5,arrowlength=5,arrowwidth=2,arrowinset=0.2](384.861,97.067)(20.435,-180,0)
    \Arc[arrow,arrowpos=0.5,arrowlength=5,arrowwidth=2,arrowinset=0.2,clock](425.732,97.067)(20.435,-180,-360)
    \Arc[arrow,arrowpos=0.5,arrowlength=5,arrowwidth=2,arrowinset=0.2](425.732,97.067)(20.435,-180,0)
    \Line[arrow,arrowpos=0.5,arrowlength=5,arrowwidth=2,arrowinset=0.2](446.167,97.067)(476.819,66.414)
    \Line[arrow,arrowpos=0.5,arrowlength=5,arrowwidth=2,arrowinset=0.2](446.167,97.067)(476.819,127.72)
    \Line[arrow,arrowpos=0.5,arrowlength=5,arrowwidth=2,arrowinset=0.2](160.075,35.761)(190.728,5.109)
    \Line[arrow,arrowpos=0.5,arrowlength=5,arrowwidth=2,arrowinset=0.2](160.075,-25.544)(190.728,5.109)
    \Arc[arrow,arrowpos=0.5,arrowlength=5,arrowwidth=2,arrowinset=0.2,clock](211.163,5.109)(20.435,-180,-360)
    \Arc[arrow,arrowpos=0.5,arrowlength=5,arrowwidth=2,arrowinset=0.2](211.163,5.109)(20.435,-180,0)
    \Arc[arrow,arrowpos=0.5,arrowlength=5,arrowwidth=2,arrowinset=0.2,clock](252.033,5.109)(20.435,-180,-360)
    \Arc[arrow,arrowpos=0.5,arrowlength=5,arrowwidth=2,arrowinset=0.2](252.033,5.109)(20.435,-180,0)
    \Arc[arrow,arrowpos=0.5,arrowlength=5,arrowwidth=2,arrowinset=0.2,clock](292.903,5.109)(20.435,-180,-360)
    \Arc[arrow,arrowpos=0.5,arrowlength=5,arrowwidth=2,arrowinset=0.2](292.903,5.109)(20.435,-180,0)
    \Line[arrow,arrowpos=0.5,arrowlength=5,arrowwidth=2,arrowinset=0.2](313.339,5.109)(343.991,35.761)
    \Line[arrow,arrowpos=0.5,arrowlength=5,arrowwidth=2,arrowinset=0.2](313.339,5.109)(343.991,-25.544)
    \Text(149.858,5.109)[lb]{\Large{\Black{$+$}}}
  \end{picture}
}
\end{center}
\caption{The exact $2\rightarrow 2$ scattering amplitude.\label{Vertex}}
\end{figure*}

There are non-zero contributions from the two-particle matrix element of $M^{ij,kl}(q)$.    We shall compute those to all orders in the coupling $g$.   In order to do so, we require the exact $2\rightarrow 2'$ scattering amplitude, which is shown in Fig.~\ref{Vertex} and given in $d$ spatial dimensions by 
\begin{equation}
\mathcal{A}^{-1}(q) = -\frac{1}{g}-\frac{m}{(4\pi)^{d/2}}\Gamma(1-d/2)\left[-m(\omega-\frac{1}{2}E_{\vec{q}}+i\epsilon)\right]^{d/2-1}.
\label{A}
\end{equation}
where $q=(\omega,{\vec q})$ is the total energy and momentum of the initial state.   In particular, for $d=3$ this gives the relation
\beq
a={mg\over 4\pi}
\eeq
between scattering length and the coupling constant $g$, indicating that in dimensional regularization, the unitary limit $a\rightarrow\infty$ corresponds to $g\rightarrow\infty$.

\begin{figure*}[t]
\begin{center}
\fcolorbox{white}{white}{
  \begin{picture}(386,108) (79,-31)
    \SetWidth{0.5}
    \SetColor{Black}
    \Text(119,-36)[lb]{\Large{\Black{$(a)$}}}
    \Text(263,-36)[lb]{\Large{\Black{$(b)$}}}
    \Text(407,-36)[lb]{\Large{\Black{$(c)$}}}
    \SetWidth{1.0}
    \Line[arrow,arrowpos=0.5,arrowlength=5,arrowwidth=2,arrowinset=0.2](80,76)(128,28)
    \Line[arrow,arrowpos=0.5,arrowlength=5,arrowwidth=2,arrowinset=0.2](80,-20)(128,28)
    \Line[arrow,arrowpos=0.5,arrowlength=5,arrowwidth=2,arrowinset=0.2](128,28)(176,76)
    \Line[arrow,arrowpos=0.5,arrowlength=5,arrowwidth=2,arrowinset=0.2](128,28)(176,-20)
    \GOval(128,28)(13,13)(0){0.882}
    \COval(160,-4)(4.243,4.243)(45.0){Black}{White}\Line(157.879,-6.121)(162.121,-1.879)\Line(157.879,-1.879)(162.121,-6.121)
    \COval(96,60)(4.243,4.243)(45.0){Black}{White}\Line(93.879,57.879)(98.121,62.121)\Line(93.879,62.121)(98.121,57.879)
    \Line[arrow,arrowpos=0.5,arrowlength=5,arrowwidth=2,arrowinset=0.2](224,76)(272,28)
    \Line[arrow,arrowpos=0.5,arrowlength=5,arrowwidth=2,arrowinset=0.2](224,-20)(272,28)
    \Line[arrow,arrowpos=0.5,arrowlength=5,arrowwidth=2,arrowinset=0.2](272,28)(320,-20)
    \Line[arrow,arrowpos=0.5,arrowlength=5,arrowwidth=2,arrowinset=0.2](272,28)(320,76)
    \GOval(272,28)(13,13)(0){0.882}
    \COval(240,60)(4.243,4.243)(45.0){Black}{White}\Line(237.879,57.879)(242.121,62.121)\Line(237.879,62.121)(242.121,57.879)
    \COval(240,-4)(4.243,4.243)(45.0){Black}{White}\Line(237.879,-6.121)(242.121,-1.879)\Line(237.879,-1.879)(242.121,-6.121)
    \Line[arrow,arrowpos=0.5,arrowlength=5,arrowwidth=2,arrowinset=0.2](368,76)(416,28)
    \Line[arrow,arrowpos=0.5,arrowlength=5,arrowwidth=2,arrowinset=0.2](368,-20)(416,28)
    \Line[arrow,arrowpos=0.5,arrowlength=5,arrowwidth=2,arrowinset=0.2](416,28)(464,-20)
    \Line[arrow,arrowpos=0.5,arrowlength=5,arrowwidth=2,arrowinset=0.2](416,28)(464,76)
    \GOval(416,28)(13,13)(0){0.882}
    \COval(384,60)(4.243,4.243)(45.0){Black}{White}\Line(381.879,57.879)(386.121,62.121)\Line(381.879,62.121)(386.121,57.879)
    \COval(400,44)(4.243,4.243)(45.0){Black}{White}\Line(397.879,41.879)(402.121,46.121)\Line(397.879,46.121)(402.121,41.879)
  \end{picture}
}
\caption{All tree graph topologies contributing to the the $2 \rightarrow 2$ matrix element of the operator $M^{ij,kl}(q)$.   The shaded blobs denote exact vertices, given in Fig.~\ref{Vertex}.  The symbol $\otimes$ denotes an insertion of the operator $\tau^{ij}$.  \label{2BodyTreeGraphs}}
\end{center}
\end{figure*}
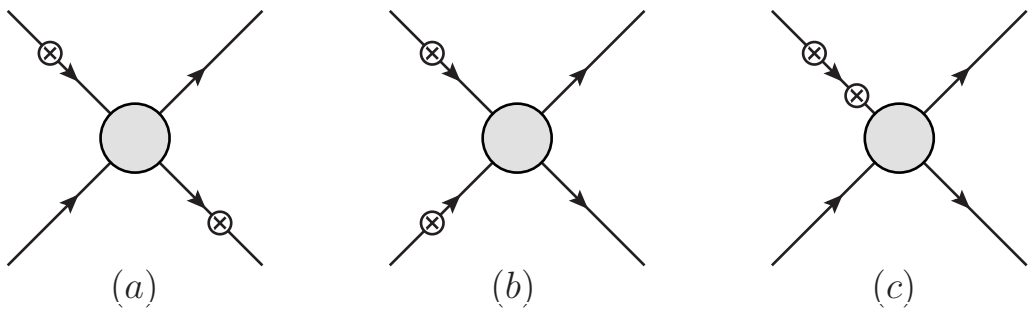

\begin{figure*}[t]
\begin{center}
\fcolorbox{white}{white}{
  \begin{picture}(451,208) (40,-36)
    \SetWidth{1.0}
    \SetColor{Black}
    \Arc[arrow,arrowpos=0.5,arrowlength=5,arrowwidth=2,arrowinset=0.2,clock](106.813,135.125)(33.033,179.256,0.744)
    \Line[arrow,arrowpos=0.5,arrowlength=5,arrowwidth=2,arrowinset=0.2](41.181,168.156)(73.783,135.554)
    \Line[arrow,arrowpos=0.5,arrowlength=5,arrowwidth=2,arrowinset=0.2](41.181,102.953)(73.783,135.554)
    \Arc[arrow,arrowpos=0.5,arrowlength=5,arrowwidth=2,arrowinset=0.2](106.813,135.983)(33.033,-179.256,-0.744)
    \GOval(73.783,135.554)(11.153,11.153)(0){0.882}
    \Line[arrow,arrowpos=0.5,arrowlength=5,arrowwidth=2,arrowinset=0.2](139.844,135.554)(172.446,168.156)
    \Line[arrow,arrowpos=0.5,arrowlength=5,arrowwidth=2,arrowinset=0.2](139.844,135.554)(172.446,102.953)
    \GOval(139.844,135.554)(11.153,11.153)(0){0.882}
    \COval(52.334,157.861)(3.64,3.64)(45.0){Black}{White}\Line(50.514,156.041)(54.154,159.681)\Line(50.514,159.681)(54.154,156.041)
    \COval(107.242,168.156)(3.64,3.64)(45.0){Black}{White}\Line(105.422,166.336)(109.062,169.976)\Line(105.422,169.976)(109.062,166.336)
    \Line[arrow,arrowpos=0.5,arrowlength=5,arrowwidth=2,arrowinset=0.2](293.415,135.554)(326.017,168.156)
    \Arc[arrow,arrowpos=0.5,arrowlength=5,arrowwidth=2,arrowinset=0.2](260.384,135.983)(33.033,-179.256,-0.744)
    \Arc[arrow,arrowpos=0.5,arrowlength=5,arrowwidth=2,arrowinset=0.2,clock](260.384,135.125)(33.033,179.256,0.744)
    \Line[arrow,arrowpos=0.5,arrowlength=5,arrowwidth=2,arrowinset=0.2](194.752,168.156)(227.354,135.554)
    \Line[arrow,arrowpos=0.5,arrowlength=5,arrowwidth=2,arrowinset=0.2](194.752,102.953)(227.354,135.554)
    \Line[arrow,arrowpos=0.5,arrowlength=5,arrowwidth=2,arrowinset=0.2](293.415,135.554)(326.017,102.953)
    \COval(238.507,157.861)(3.64,3.64)(45.0){Black}{White}\Line(236.687,156.041)(240.327,159.681)\Line(236.687,159.681)(240.327,156.041)
    \COval(282.262,157.861)(3.64,3.64)(45.0){Black}{White}\Line(280.442,156.041)(284.082,159.681)\Line(280.442,159.681)(284.082,156.041)
    \GOval(227.354,135.554)(11.153,11.153)(0){0.882}
    \GOval(293.415,135.554)(11.153,11.153)(0){0.882}
    \Arc[arrow,arrowpos=0.5,arrowlength=5,arrowwidth=2,arrowinset=0.2,clock](425.108,135.125)(33.033,179.256,0.744)
    \Arc[arrow,arrowpos=0.5,arrowlength=5,arrowwidth=2,arrowinset=0.2](425.108,135.983)(33.033,-179.256,-0.744)
    \Line[arrow,arrowpos=0.5,arrowlength=5,arrowwidth=2,arrowinset=0.2](359.476,168.156)(392.078,135.554)
    \Line[arrow,arrowpos=0.5,arrowlength=5,arrowwidth=2,arrowinset=0.2](359.476,102.953)(392.078,135.554)
    \Line[arrow,arrowpos=0.5,arrowlength=5,arrowwidth=2,arrowinset=0.2](458.139,135.554)(490.741,168.156)
    \Line[arrow,arrowpos=0.5,arrowlength=5,arrowwidth=2,arrowinset=0.2](458.139,135.554)(490.741,102.953)
    \COval(425.537,168.156)(3.64,3.64)(45.0){Black}{White}\Line(423.717,166.336)(427.357,169.976)\Line(423.717,169.976)(427.357,166.336)
    \COval(425.537,102.953)(3.64,3.64)(45.0){Black}{White}\Line(423.717,101.133)(427.357,104.773)\Line(423.717,104.773)(427.357,101.133)
    \GOval(392.078,135.554)(11.153,11.153)(0){0.882}
    \GOval(458.139,135.554)(11.153,11.153)(0){0.882}
    \Arc[arrow,arrowpos=0.5,arrowlength=5,arrowwidth=2,arrowinset=0.2,clock](227.783,15.014)(33.033,-179.256,-360.744)
    \Arc[arrow,arrowpos=0.5,arrowlength=5,arrowwidth=2,arrowinset=0.2,clock](293.844,15.014)(33.033,-179.256,-360.744)
    \Arc[arrow,arrowpos=0.5,arrowlength=5,arrowwidth=2,arrowinset=0.2](227.783,15.014)(33.033,-179.256,-0.744)
    \Line[arrow,arrowpos=0.5,arrowlength=5,arrowwidth=2,arrowinset=0.2](326.875,14.585)(359.476,47.187)
    \Line[arrow,arrowpos=0.5,arrowlength=5,arrowwidth=2,arrowinset=0.2](326.875,14.585)(359.476,-18.017)
    \Line[arrow,arrowpos=0.5,arrowlength=5,arrowwidth=2,arrowinset=0.2](162.15,48.045)(194.752,15.443)
    \Line[arrow,arrowpos=0.5,arrowlength=5,arrowwidth=2,arrowinset=0.2](162.15,-18.017)(194.752,14.585)
    \Arc[arrow,arrowpos=0.5,arrowlength=5,arrowwidth=2,arrowinset=0.2](293.844,15.014)(33.033,-179.256,-0.744)
    \COval(227.354,48.045)(3.64,3.64)(45.0){Black}{White}\Line(225.534,46.225)(229.174,49.865)\Line(225.534,49.865)(229.174,46.225)
    \COval(293.415,48.045)(3.64,3.64)(45.0){Black}{White}\Line(291.595,46.225)(295.235,49.865)\Line(291.595,49.865)(295.235,46.225)
    \GOval(260.813,14.585)(11.153,11.153)(0){0.882}
    \GOval(193.036,17.159)(11.153,11.153)(0){0.882}
    \GOval(326.875,14.585)(11.153,11.153)(0){0.882}
    \Text(99.242,80.646)[lb]{\Large{\Black{$(a)$}}}
    \Text(252.813,80.646)[lb]{\Large{\Black{$(b)$}}}
    \Text(417.537,80.646)[lb]{\Large{\Black{$(c)$}}}
    \Text(252.813,-40.323)[lb]{\Large{\Black{$(d)$}}}
  \end{picture}
}
\caption{All loop graph topologies contributing to the the $2 \rightarrow 2$ matrix element of the operator $M^{ij,kl}(q)$.   The shaded blobs denote exact vertices, given in Fig.~\ref{Vertex}.  The symbol $\otimes$ denotes an insertion of the operator $\tau^{ij}$.   \label{2BodyLoopGraphs}}
\end{center}
\end{figure*}
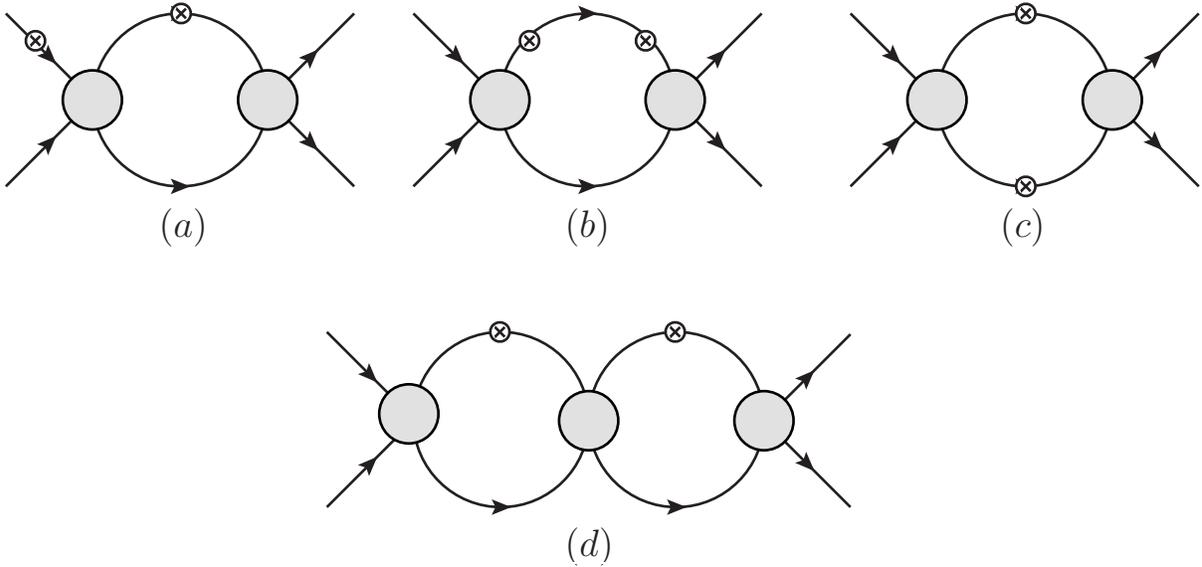

The full set of diagrams contributing to the $2\rightarrow 2'$ matrix element are given in Fig.~\ref{2BodyTreeGraphs} and Fig.~\ref{2BodyLoopGraphs}.   In the limit of zero external momentum and ${\vec q}\rightarrow 0$, the diagrams with operator insertions on external lines (all graphs in Fig.~\ref{2BodyTreeGraphs}, and graphs (a) in Fig.~\ref{2BodyLoopGraphs}) vanish identically.   It follows from Eq.~(\ref{eq:trule}) that each loop integrand in Fig.~\ref{2BodyLoopGraphs}(d) is proportional to ${\vec \ell}_i {\vec \ell}_j - {\vec\ell^2\over d}\delta_{ij}$ (where ${\vec \ell}$ is loop momentum).    By rotational invariance, the angular integrals give ${\vec l}_i {\vec l}_j\rightarrow \delta_{ij} {\vec \ell^2}/d$ and therefore Fig.~\ref{2BodyLoopGraphs}(d) vanishes.

The remaining diagrams are, after performing the energy integrals by residues,
\begin{equation}
\mbox{Fig.~\ref{2BodyLoopGraphs}(b)} = {i{\cal A}(0)^2\over 2m^2} \int {d^d {\vec\ell}\over (2\pi)^d} {1\over E_{\vec l}^2 }{1\over \omega - 2 E_{\vec l} + i\epsilon}{\left( {\vec \ell}_i {\vec \ell}_j - {\vec\ell^2\over d}\delta_{ij}\right) \left( {\vec \ell}_k {\vec \ell}_l - {\vec\ell^2\over d}\delta_{kl}\right)} + (\omega\rightarrow -\omega),
\end{equation}
and 
\begin{equation}
\mbox{Fig.~\ref{2BodyLoopGraphs}(c)} = {i{\cal A}(0)^2\over m^2\omega} \int {d^d {\vec\ell}\over (2\pi)^d} {1\over E_{\vec l} }{1\over \omega - 2 E_{\vec l} + i\epsilon}{\left( {\vec \ell}_i {\vec \ell}_j - {\vec\ell^2\over d}\delta_{ij}\right) \left( {\vec \ell}_k {\vec \ell}_l - {\vec\ell^2\over d}\delta_{kl}\right)} + (\omega\rightarrow -\omega),
\end{equation}
where $E_{\vec l} = {\vec l}^2/2m$.   Using rotational invariance, we make the replacements
\begin{eqnarray}
{\vec l}_i {\vec l}_j &\rightarrow& {{\vec l}^2\over d} \delta_{ij} \\
{\vec l}_i {\vec l}_j  {\vec l}_k {\vec l}_l  &\rightarrow& {{\vec l}^4\over d(d+2)} \left[\delta_{ij} \delta_{kl}+ \delta_{ik} \delta_{jl}+\delta_{il} \delta_{kj}\right],
\end{eqnarray}
and use
\beq
\int {d^d {\vec\ell}\over (2\pi)^d} {E^n_{\vec l}\over \omega - 2 E_{\vec l} + i\epsilon} = -{1\over 2} \left({1\over 2m}\right)^{n-1} {\Gamma(1-n-d/2)\over (4\pi)^{d/2}} {\Gamma(d/2+n)\over \Gamma(d/2)} \left[-m(\omega+i\epsilon)\right]^{d/2+n-1}.
\label{eq:easyI}
\eeq
This gives the exact $2\rightarrow 2'$ matrix element at zero external momentum and ${\vec q}=0$:
\begin{equation}
\langle 2'|M^{ij,kl}(q)|2\rangle= -{4 m i {{\cal A}(0)}^2\over d (d+2)} {\Gamma(1-d/2)\over (4\pi)^{d/2}}\left[\left(-m\omega-i\epsilon\right)^{d/2-1} +\left(m\omega-i\epsilon\right)^{d/2-1} \right]\cdot\left(\delta_{ik} \delta_{jl}+\delta_{il} \delta_{kj}-{2\over d} \delta_{ij} \delta_{kl}\right),
\end{equation}
which indicates that the OPE for $M^{ij,kl}(q\rightarrow\infty)$ contains the four-fermion contact operator ${\cal O}_4 = \psi_1^\dagger\psi_2^\dagger \psi_1\psi_2$.    Using $\langle 2' | \mathcal{O}_4(0) | 2\rangle = -\mathcal{A}(0)^2/g^2$ (see~\cite{GnR}), we obtain
\beq
\label{eq:MOPE}
M^{xy,xy}(\omega\rightarrow\infty,{\vec q}=0) \sim  {4 m i \over d (d+2)} {\Gamma(1-d/2)\over (4\pi)^{d/2}}\left[\left(-m\omega-i\epsilon\right)^{d/2-1} +\left(m\omega-i\epsilon\right)^{d/2-1} \right] \cdot\left[g^2 {\cal O}_4(0)\right]+\cdots,
\eeq
where $\ldots$ denotes the contribution to the OPE from operators of higher dimension.  

\subsection{\label{Results} Asymptotics and Sum Rules}

The results of the previous section can be combined with those of ref.~\cite{GnR} on the dynamical structure function $S(\omega,{\vec q})$ to obtain the asymptotic behavior of the frequency dependent viscosities.   We will focus on the case $d=3$.      By Eq.~(\ref{Kubo3}), and Eq.~(\ref{eq:MOPE}) we obtain
\begin{equation}
\mathrm{Re}\,\eta(\omega) \underset{\omega \to \infty}{\sim}\frac{m^2}{15\pi} (m\omega)^{-1/2}\cdot  \mathcal{C}(a) + {\cal O}\left({\mu\over\omega}\right)^{1}.
\label{Asymptotic3d}
\end{equation}
where we have used the relation $\mbox{Im} \chi^{xy,xy}(\omega,{\vec q})_R = \tanh(\omega/2T)\cdot \mathrm{Re}\, \langle M^{xy,xy}(\omega,{\vec q})\rangle$ valid for real $\omega>0$.   In Eq.~(\ref{Asymptotic3d}), ${\cal C}(a)$ is the by now ubiquitous Tan contact~\cite{tan1,tan2,tan3}, which is the expectation value $\mathcal{C}(a)=\left\langle -g^2\mathcal{O}_4\right\rangle$ and is proportional to the thermodynamic quantity $\partial {\cal F}/\partial a^{-1}$~\cite{Braaten:2010if}.    

The fact that $\mathrm{Re}\,\eta(\omega)\sim \omega^{-1/2}$ at large frequencies is a simple consequence of Eq.~(\ref{eq:GOPE}), given that the scaling dimensions are $\Delta(\tau_{ij})=5$ and $\Delta(g^2{\cal O}_4)=4$.   The numerical coefficient is in agreement with the result of~\cite{EHZ}.  However, it follows from our analysis that Eq.~(\ref{Asymptotic3d}) is exact to all orders in the coupling constant $g$, and consequently gives the \emph{exact} dependence on the parameter $a\sqrt{m\omega}$, for arbitrary values of the scattering length $a$.   There are corrections to Eq.~(\ref{Asymptotic3d}) from operators starting at dimension $\Delta=6$ in the OPE (for bosons, there would be corrections from three-body contact operators at $\Delta=5$~\cite{braaten2011}).   Such terms are suppressed by a factor of $(\mu/\omega)^{1}$ in the large frequency limit.

Using Eq.~(\ref{eq:K5}), Eq.~(\ref{Asymptotic3d}) and Eq.~(\ref{eq:recc}) from appendix~\ref{ope} (reproduced from ref.~\cite{GnR}) we also get the asymptotic bulk viscosity,
\begin{equation}
\mathrm{Re}\,\zeta(\omega) \underset{\omega \to \infty}{\sim} \frac{m^2}{36\pi} \cdot { (m\omega)^{-1/2}\over 1 + (a\sqrt{m\omega})^2}\cdot {\cal C}(a)+   {\cal O}\left({\mu\over\omega}\right)^{1},
\end{equation}
which vanishes in the limit $a\rightarrow\infty$, as expected by Schrodinger invariance~\cite{Nishida:2007pj}.   

From Eq.~\eqref{Kubo3}, and Eq.~\eqref{eq:MOPE} one gets the following sum rule for $\mathrm{Re}\,\eta(\omega)$: 
\begin{equation}
\frac{1}{\omega_0^2}\int_0^\infty \frac{\mathrm{d}\omega^2}{\pi} e^{-\omega^2/\omega_0^2} \omega \mathrm{Re}\,\eta(\omega) = {m\over 60\pi^2} \Gamma(1/4) (m\omega_0)^{1/2} \cdot {\cal C}(a) + {\cal O}\left({\mu\over\omega}\right)^{1},
\label{SumRule3}
\end{equation}
which exhibits the exact functional dependence on the parameter $1/a\sqrt{m\omega_0}$ at this order in the OPE. From Eq.~(\ref{eq:K5}) and Eq.~(\ref{eq:Cc}) we also obtain a bulk viscosity sum rule
\begin{align}
\label{eq:zsr}
\frac{1}{\omega_0^2}\int_0^\infty \frac{\mathrm{d}\omega^2}{\pi} e^{-\omega^2/\omega_0^2} \omega \mathrm{Re}\,\zeta(\omega) &=  {\sqrt{2}m\over 36\pi} (m\omega_0)^{1/2} F(a\sqrt{m\omega_0}) \cdot {\cal C}(a) + {\cal O}\left({\mu\over\omega}\right)^{1} \notag \\
&= \frac{\sqrt{2}m}{36\pi\Gamma(1/4)}\left(m\omega_0\right)^{1/2}\left(a\sqrt{m\omega_0}\right)^{-2} \times \notag \\
&\hspace{5mm} \left( 1 - \frac{\sqrt{2}}{2\pi}\Gamma(1/4)^2\left(a\sqrt{m\omega_0}\right)^{-2} - \sqrt{2}\Gamma(1/4)\left(a\sqrt{m\omega_0}\right)^{-3} + ...\right) \cdot {\cal C}(a) + {\cal O}\left({\mu\over\omega}\right)^{1}, 
\end{align}
where the function $F(a\sqrt{m\omega_0})$ is obtained by Borel resumming the Wilson coefficient in Eq.~(\ref{eq:Cc}).  It is given by 
\begin{equation}
F(a\sqrt{m\omega_0}) = \left. z^{5/4} \frac{\mathrm{d}^2}{\mathrm{d}z^2} g(z) \right|_{z=(a\sqrt{m\omega_0})^{-4}},
\end{equation}
where
\begin{equation}
g(z) = \sqrt{2}\left( 1-e^{-z} \right) + z^{5/4}\frac{4}{\Gamma(1/4)}e^{-z}{\tilde\gamma}(5/4,-z) - z^{7/4}\frac{2\sqrt{2}}{3\pi}\Gamma(1/4) e^{-z}{\tilde\gamma}(7/4,-z), 
\end{equation}
and ${\tilde\gamma}(s,z)$ is defined by the analytic continuation of the integral
\begin{equation}
{\tilde\gamma}(s,z) = z^{-s} \int_0^z dt \, t^{s-1} e^{-t}, \hspace{5mm} (\mathrm{Re}\, z > 0).
\end{equation}
The function $\tilde\gamma(s,z)$ is proportional to $e^{-z}$ times the confluent hypergeometric function ${}_1 F_1(1,s,z)$ and is therefore an entire function over the complex $z$-plane for fixed $s$. Both sum rules receive corrections from operators at dimension $\Delta=6$.
%
%

\section{\label{Conclusion} Conclusion}
In this paper we have obtained new Borel sum rules for the spectral shear and bulk viscosity of a non-relativistic Fermi gas.   By adjusting the Borel parameter $\omega_0$, the sum rules can be used to constrain the long time scale transport behavior of these systems.   Our results are exact in $a$, but receive corrections of relative size $({\mu/\omega_0})^1\ll 1$ from condensates of dimension $\Delta>5$.    

We find it particularly striking that the scattering length dependence of our shear viscosity sum rule is given \emph{exactly} by Tan's function ${\cal C}(a)$.   Thus, at least for the particular case of the shear viscosity, our results provide a simple direct link between transport and the theoretically more tractable thermodynamic behavior of this strongly coupled system.

\section{Acknowledgments}

We thank I. Rothstein for comments on the manuscript.   This work is supported in part by DOE grant DE-FG-02-92ER40704 and by an OJI award.
\\
\\
{\bf Note added}:   While this paper was being completed, ref.~\cite{hofmann} appeared on the arXiv which has some overlap with the results presented here.

\appendix

\section{Current-Current OPE}
\label{ope}

In this appendix we give results for the operator product expansion of 
\begin{equation}
M^{i,j}(q) \equiv \int \mathrm{d}^{d+1}x \, e^{iq\cdot x} \, T\left\{J^i(x)J^j(0)\right\},
\end{equation}
in the limit $\omega\rightarrow\infty$ but arbitrary ${\vec q}^2/2m\omega$.   Our results are valid up to corrections from operators of dimension $\Delta>5$.   For terms up to operator dimension $\Delta=5$ we find exact Wilson coefficients to all orders in the coupling constant $g$.   

We take as our starting point the OPE for the density correlator,
\begin{equation}
M(q) \equiv \int \mathrm{d}^{d+1}x \, e^{iq\cdot x} \, T\left\{n(x)n(0)\right\},
\end{equation}
which was computed including the contributions of operators $\Delta\leq 5$ in ref.~\cite{GnR}.   As we use that result in the determination of the $M^{i,j}(q)$ OPE, we now recall them.    Matching in the one-particle sector generates  the operators: 
\begin{equation*}
n, \hspace{5mm} \partial_i n, \hspace{5mm} J_i, \hspace{5mm} \partial_i\partial_j n, \hspace{5mm} (\partial_i J_j + \partial_j J_i),
\end{equation*}
\begin{equation}
\frac{1}{2m}\psi^\dagger \overleftrightarrow{\partial_i} \overleftrightarrow{\partial_j} \psi, \hspace{5mm} \psi^\dagger \left(i\partial_t + \frac{\nabla^2}{2m}\right)\psi, \hspace{5mm} \psi^\dagger \overleftarrow{\left(-i\partial_t + \frac{\nabla^2}{2m}\right)}\psi,
\label{operators}
\end{equation}
in the OPE of $M(q)$,  while matching in the two-particle sector gives rise to the additional operator
\begin{equation}
\mathcal{O}_4 = \psi_1^\dagger\psi_2^\dagger\psi_1\psi_2.
\end{equation}
The expression for the OPE simplifies if one restricts attention to matrix elements in a translationally and rotationally invariant many-body state $|\Omega\rangle$. In that case, excluding $n$, the VEVs of the operators appearing in the first row of Eq.\eqref{operators} vanish.   The result is then ($\langle\cdots\rangle$ denotes a matrix element in the state $|\Omega\rangle$, or a thermodynamic average at finite temperature),
\begin{equation}
\left\langle M(q) \right\rangle = C_n(q) \left\langle n \right\rangle + C_{\mathcal{H}}(q) \left\langle \mathcal{H} \right\rangle + C_c(q) \left\langle g^2 \mathcal{O}_4 \right\rangle,
\end{equation}
where 
\begin{equation}
\mathcal{H} = \frac{1}{2m}\left|\nabla\psi\right|^2 - g\psi_1^\dagger\psi_2^\dagger\psi_1\psi_2,
\end{equation}
\begin{equation}
C_n(q) = \frac{2iE_{\vec{q}}}{(\omega-E_{\vec{q}})(\omega+E_{\vec{q}})}
\end{equation}
\begin{equation}
C_{\mathcal{H}}(q) = \frac{4i}{d}E_{\vec{q}}\left[ \frac{1}{(\omega-E_{\vec{q}})^3}-\frac{1}{(\omega+E_{\vec{q}})^3}\right]
\end{equation}
\begin{align}
C_c(q) = &i\left[\mathcal{A}(q)\left(I_1(q) - \frac{2}{g}\frac{1}{\omega-E_{\vec{q}}}\right)^2 - \frac{1}{2}I_2(q) - \frac{1}{\omega}I_1(q) + (q \rightarrow -q)\right] \notag \\
&+ \frac{2i}{g}\left[\frac{1}{\omega-E_{\vec{q}}} + \frac{1}{\omega+E_{\vec{q}}}\right]^2 + \frac{4i}{g}\frac{E_{\vec{q}}}{d}\left[\frac{1}{(\omega-E_{\vec{q}})^3} - \frac{1}{(\omega+E_{\vec{q}})^3}\right],
\end{align}
and
\begin{align}
I_\alpha (q) &\equiv \int \frac{\mathrm{d}^d\vec{\ell}}{(2\pi)^d} \left( \frac{1}{E_{\vec{\ell}}} \right)^\alpha \frac{1}{\omega-E_{\vec{\ell}}-E_{\vec{\ell}+\vec{q}}+i\epsilon} \notag \\
&= -2^\alpha \left(\frac{m}{4\pi}\right)^{\frac{d}{2}} \frac{\Gamma(\alpha+1-\frac{d}{2})\Gamma(\frac{d}{2}-\alpha)}{\Gamma(\frac{d}{2})} \left( \frac{1}{\frac{1}{2}E_{\vec{q}} - \omega - i\epsilon} \right)^{1+\alpha-\frac{d}{2}} \times \notag \\ 
&\hspace{7mm} {}_2F_1\left(\alpha,\alpha+1-\frac{d}{2},\frac{d}{2},\frac{\frac{1}{2}E_{\vec{q}}}{\omega-\frac{1}{2}E_{\vec{q}}+i\epsilon}\right).
\end{align}
Note that
\begin{equation}
I_0(q) = \frac{1}{g} + \frac{1}{\mathcal{A}(q)},
\end{equation}
where $\mathcal{A}(q)$ is the full $2\to 2$ scattering amplitude from Eq.~(\ref{A}), and that the integral in Eq.~(\ref{eq:easyI}) corresponds to the case $I_{-n}(\omega,{\vec q}=0)$.

In order to compute $M^{i,j}(q)$, we have found it convenient to use Ward identities (current conservation) to relate it to the OPE of $M(q)$ given above and to the OPE with one insertion of the current ${\vec J}$ and one insertion of the density operator $n$.    We only display results for the expectation value $\langle  M^{i,j}(q)\rangle$ in a state that is invariant under rotations and translation, as the full operator relation is not useful for obtaining the physical observables discussed in sec.~\ref{Results}.   In this case, we find
\begin{equation}
\left\langle M^{i,j}(q) \right\rangle = \frac{q^i q^j}{\vec{q}^2} L(q)  + \left(\delta^{ij}-\frac{q^i q^j}{\vec{q}^2}\right) T(q),
\end{equation}
where
\begin{align}
L(q) &= C^{(1)}_n(q) \left\langle n \right\rangle + C^{(1)}_{\mathcal{H}}(q) \left\langle \mathcal{H} \right\rangle + C^{(1)}_c(q) \left\langle g^2 \mathcal{O}_4 \right\rangle, \\
T(q) &= C^{(2)}_{\mathcal{H}}(q) \left\langle \mathcal{H} \right\rangle + C^{(2)}_c(q) \left\langle g^2 \mathcal{O}_4 \right\rangle,
\end{align}
and
\begin{align}
C^{(1)}_n(q) &= \frac{\omega^2}{\vec{q}^2} C_n(q) - \frac{i}{m} ,\\
C^{(1)}_{\mathcal{H}}(q) &= \frac{\omega^2}{\vec{q}^2} C_{\mathcal{H}}(q), \\
C^{(1)}_c(q) &= \frac{\omega^2}{\vec{q}^2} C_c(q), \\
C^{(2)}_{\mathcal{H}}(q) &= \frac{4i}{d} \frac{E_{\vec{q}}}{m(\omega-E_{\vec{q}})(\omega+E_{\vec{q}})},
\end{align}
\begin{align}
C^{(2)}_c(q) = \frac{i}{m^2}&\left[ -\frac{2m}{(d-1)\omega E_{\vec{q}}} I_{-1}(q) + \frac{3m}{(d-1) E_{\vec{q}}}I_0(q) - \frac{m\left(E_{\vec{q}}^2 - 2\omega E_{\vec{q}} + 3\omega^2\right)}{2(d-1)\omega E_{\vec{q}}}I_1(q)\right. \notag \\
&+ \left.\frac{m\left(\omega-E_{\vec{q}}\right)^2}{4(d-1) E_{\vec{q}}}I_2(q) + (q\rightarrow -q)\right] + \frac{4i}{g}\frac{E_{\vec{q}}}{d}\frac{1}{m(\omega-E_{\vec{q}})(\omega+E_{\vec{q}})}.
\end{align}
The causal and time-ordered Green's functions for the current were defined in Eq.~(\ref{eq:chiR}) and Eq.~(\ref{eq:GOPE}), respectively. Since $G^{i,j}(q) = iM^{i,j}(q)$ and $\mathrm{Im}\,G^{i,j}(q) = \mathrm{Im}\,\chi_R^{i,j}(q)$ (for real $\omega >0$), we have for $q^\mu \rightarrow\infty$
\begin{align}
\mathrm{Im}\, \chi_T(q)  &= \mathrm{Re}\, T(q)  =  \left\langle g^2\mathcal{O}_4 \right\rangle \mathrm{Re}\, C_c^{(2)}(q), \\
\mathrm{Im} \,\chi_L(q) &= \mathrm{Re} \, L(q)  = \left\langle g^2\mathcal{O}_4 \right\rangle \mathrm{Re}\, C_c^{(1)}(q) = \frac{\omega^2}{\vec{q}^2} \left\langle g^2\mathcal{O}_4 \right\rangle \mathrm{Re}\, C_c(q).
\end{align}

In the limit ${\vec q}\rightarrow 0$ the relevant Wilson coefficients have the expansion, for $\omega>0$,
\begin{equation}
\mathrm{Re}\, C_c^{(2)}(q) = \frac{\vec{q}^2}{(m\omega)^2}\cdot \frac{4}{d(d+2)} \mathrm{Im} \left[{\cal A}(\omega)^{-1} + {\cal A}(-\omega)^{-1}\right] + \mathcal{O}\left(\vec{q}^4\right),
\end{equation}
%
%
\begin{eqnarray}
\nn
\mathrm{Re}\, C_c(q) = \left({\vec{q}^2 \over m\omega}\right)^2 {1\over d^2 (d+2)\omega^2}\mathrm{Im}\left[{1\over {\cal A}(\omega)}\left( 8(d-1) - (d-2)^2 (d+2) \left({{\cal A}(\omega)\over g}\right)^2\right) +  (\omega \rightarrow -\omega)\right]+ \mathcal{O}\left(\vec{q}^6\right),\\
\end{eqnarray}
where ${\cal A}(\omega)={\cal A}(\omega,{\vec q}=0)$ is the $2\rightarrow 2'$ amplitude of Eq.~(\ref{A}) at center-of-mass energy $\omega$.  In $d=3$ spatial dimensions, these expressions become
%
\begin{equation}
\mathrm{Re}\, C_c^{(2)}(q) = -{m\over 15\pi} \left({\vec{q}^2\over m\omega}\right) (m\omega)^{-1/2} + \mathcal{O}\left(\vec{q}^4\right),
\end{equation}
and
\begin{equation}
\label{eq:recc}
\mathrm{Re}\, C_c(q) = -{4 m^3\over 45\pi} \left({\vec{q}^2\over m\omega}\right)^2 (m\omega)^{-3/2} \cdot\left[1+{5\over 16\left(1 + (a\sqrt{m\omega})^{2}\right)}\right] + \mathcal{O}\left(\vec{q}^6\right).
\end{equation}
%
The Borel sum rules are related to the Wilson coefficients at imaginary frequency.   In $d=3$ we obtain for $\omega_0>0,$
%
\begin{equation}
C_c^{(2)}(i\omega_0) = {im\sqrt{2}\over 15\pi} \left({\vec{q}^2\over m\omega_0}\right) (m\omega_0)^{-1/2}\left[1- {5\sqrt{2}\over 4} (a\sqrt{m\omega_0})^{-1}\right] + \mathcal{O}\left(\vec{q}^4\right),
\end{equation} 
\begin{equation}
\label{eq:Cc}
C_c(i\omega_0) = -{7\sqrt{2} i m^3\over 60\pi} \left({\vec{q}^2\over m\omega_0}\right)^2 (m\omega_0)^{-3/2} \cdot\left[ 1 - {5\over 21\left(1 -\sqrt{2} (a\sqrt{m\omega_0})^{-1} + (a\sqrt{m\omega_0})^{-2}\right)}\right] + {\cal O}\left(\vec{q}^6\right),
\end{equation}

%



\end{document}